\documentclass[conference]{IEEEtran}
\IEEEoverridecommandlockouts
\usepackage{cite}
\usepackage{amsmath,amssymb,amsfonts}
\usepackage{graphicx}
\usepackage{textcomp}
\usepackage{xcolor}
\usepackage{multirow}
\usepackage{makecell}
\usepackage[draft=true]{hyperref}
\usepackage{pifont}
\usepackage{algorithm}
\usepackage{algpseudocode}
\usepackage[utf8]{inputenc}
\usepackage[english]{babel}
\usepackage{lastpage}
\usepackage{adjustbox}
\usepackage{fancyhdr}
\usepackage{balance}
\usepackage{booktabs} 
\usepackage[left=2cm, right=2cm, top=0.75in, bottom=1.2in]{geometry}
\makeatletter
\renewcommand{\fnum@figure}{Fig. \thefigure}
\makeatother

\begin{document}

\title{
Quantum-Inspired Genetic Algorithm for Robust Source Separation in Smart City Acoustics
\\
}

\author{
        \IEEEauthorblockN{
        Minh K. Quan\IEEEauthorrefmark{1},
        Mayuri Wijayasundara\IEEEauthorrefmark{1}, Sujeeva Setunge\IEEEauthorrefmark{4},
        Pubudu N. Pathirana\IEEEauthorrefmark{1}
	}
	\IEEEauthorblockA{\IEEEauthorrefmark{1}School of Engineering, Deakin University, Australia \\
	 \IEEEauthorrefmark{4}School of Engineering, Royal Melbourne Institute of Technology University, Melbourne, VIC 3000, Australia
	}}
	\markboth{}%
	{}

\maketitle
\pagenumbering{gobble} 
\begin{abstract}
The cacophony of urban sounds presents a significant challenge for smart city applications that rely on accurate acoustic scene analysis. Effectively analyzing these complex soundscapes, often characterized by overlapping sound sources, diverse acoustic events, and unpredictable noise levels, requires precise source separation. This task becomes more complicated when only limited training data is available. This paper introduces a novel Quantum-Inspired Genetic Algorithm (p-QIGA) for source separation, drawing inspiration from quantum information theory to enhance acoustic scene analysis in smart cities. By leveraging quantum superposition for efficient solution space exploration and entanglement to handle correlated sources, p-QIGA achieves robust separation even with limited data. These quantum-inspired concepts are integrated into a genetic algorithm framework to optimize source separation parameters. The effectiveness of our approach is demonstrated on two datasets: the TAU Urban Acoustic Scenes 2020 Mobile dataset, representing typical urban soundscapes, and the Silent Cities dataset, capturing quieter urban environments during the COVID-19 pandemic.  Experimental results show that the p-QIGA achieves accuracy comparable to state-of-the-art methods while exhibiting superior resilience to noise and limited training data, achieving up to 8.2 dB signal-to-distortion ratio (SDR) in noisy environments and outperforming baseline methods by up to 2 dB with only 10\% of the training data. This research highlights the potential of p-QIGA to advance acoustic signal processing in smart cities, particularly for noise pollution monitoring and acoustic surveillance.
\end{abstract}

\begin{IEEEkeywords}
Acoustic signal processing, Quantum-inspired algorithms, Smart cities, Source separation
\end{IEEEkeywords}

\section{Introduction}
\label{Section:Introduction}

The rise of smart cities has brought an abundance of opportunities to improve urban living through data-driven solutions. Acoustic scene analysis (ASA) plays a crucial role in this vision, enabling a deeper understanding of the urban environment through the analysis of sound \cite{abesser2020review}. However, the complexity of urban soundscapes, characterized by overlapping sound sources, diverse acoustic events, and unpredictable noise levels, presents a significant challenge for ASA, particularly in source separation. Traditional source separation methods, such as Independent Component Analysis (ICA) and Non-negative Matrix Factorization (NMF), often struggle with these complex soundscapes due to their limitations in handling correlated sources and their sensitivity to noise \cite{ansari2023survey}. This necessitates the exploration of more robust and adaptable techniques that can effectively disentangle individual sound sources in real-world urban scenarios.

Quantum information theory, with principles like superposition and entanglement, offers a powerful framework for signal processing and communication, enabling efficient representation of high-dimensional signals and encoding of correlated information beyond classical capabilities. Quantum-inspired genetic algorithms have been previously explored for various optimization problems. For instance, Roy et al. \cite{roy2014optimization} demonstrated the effectiveness of quantum-based genetic algorithms for optimizing complex functions. Our work builds upon these prior contributions by specifically applying this approach to the noise separation problem in complex urban acoustic scenes, with a focus on achieving robust and accurate source separation even with limited training data. However, limitations in current quantum hardware hinder the direct application of quantum algorithms to ASA. This motivates exploring quantum-inspired algorithms, adapting quantum mechanics principles to classical computation for near-term applications while leveraging the unique advantages of quantum phenomena.


















\subsection{Background and Related Work}

ASA involves tasks such as acoustic scene classification and sound event detection, with source separation being a crucial component \cite{abesser2020review}. However, traditional source separation methods like ICA \cite{xie2019investigation} and NMF \cite{bisot2016acoustic}, as well as more recent techniques like Sparse Component Analysis (SCA) \cite{asaei2016computational} and deep learning (DL) (e.g., \cite{xie2022deep, arniriparian2018fusion}), face challenges in handling correlated sources, noise robustness, computational efficiency, and data efficiency in complex urban acoustic scenes. While promising for signal processing, the direct application of quantum algorithms to ASA, particularly source separation, remains largely unexplored. Quantum algorithms, such as Quantum Principal Component Analysis (QPCA) \cite{lloyd2014quantum} and Quantum Fourier Transform (QFT) \cite{zhou2017quantum}, offer potential advantages for signal processing tasks due to their inherent ability to handle high-dimensional spaces and exploit quantum phenomena like superposition and entanglement. However, their application to source separation in complex acoustic environments requires further investigation. Similarly, Quantum Support Vector Machines (QSVM) \cite{rebentrost2014quantum} is more suited for classification tasks, while Quantum Annealing (QA) \cite{chancellor2017modernizing} shows potential for optimization but needs further evaluation in complex scenarios.

\subsection{Motivations and Key Contributions}

Our proposed Quantum-Inspired Genetic Algorithm (p-QIGA) addresses the challenge of source separation in complex urban soundscapes by incorporating quantum concepts into a genetic optimization framework. This approach allows us to effectively disentangle individual sound sources in real-world urban scenarios, even in the presence of noise and limited training data. The p-QIGA's ability to accurately separate sources enhances the identification and tracking of vehicles in noisy environments and contributes to improved urban planning by providing insights into the acoustic characteristics of different urban spaces. This research makes the following key contributions:

\begin{itemize}
\item We introduce a novel p-QIGA leveraging quantum concepts to optimize source separation in complex urban acoustic scenes, enhancing performance and robustness even with limited data.
\item Our p-QIGA effectively addresses key challenges in smart city source separation, including handling correlated sources and diverse sound events.
\item By achieving accurate and robust source separation, our p-QIGA improves performance in critical smart city applications like noise pollution monitoring and acoustic surveillance.
\end{itemize}

\section{Proposed Methodology}

\subsection{Problem Formulation: Source Separation in ASA}
Source separation in ASA aims to decompose an observed audio signal $x(t)$ into its constituent sound sources $s_i(t)$. The mixing process can be modeled as a convolutive mixture in the time domain:
\begin{equation}
x(t) = \sum_{i=1}^{N} \sum_{k=0}^{K-1} a_i(k) s_i(t-k) + n(t),
\end{equation}
or equivalently in the frequency domain:
\begin{equation}
X(f) = \sum_{i=1}^{N} A_i(f) S_i(f) + N(f),
\end{equation}
where $a_i(k)$ and $A_i(f)$ are the mixing filters, and $n(t)$ and $N(f)$ represent noise.  The goal is to estimate $s_i(t)$ (or $S_i(f)$) given $x(t)$ (or $X(f)$).  This is challenging due to factors like unknown mixing filters, correlated sources, diverse sound events, noise, reverberation, and limited training data, motivating the exploration of novel approaches like the quantum-inspired genetic algorithm proposed in this paper.

\subsection{Quantum Encoding of Acoustic Features}
\begin{figure}[htbp]
\centerline{\includegraphics[width=0.98\linewidth]{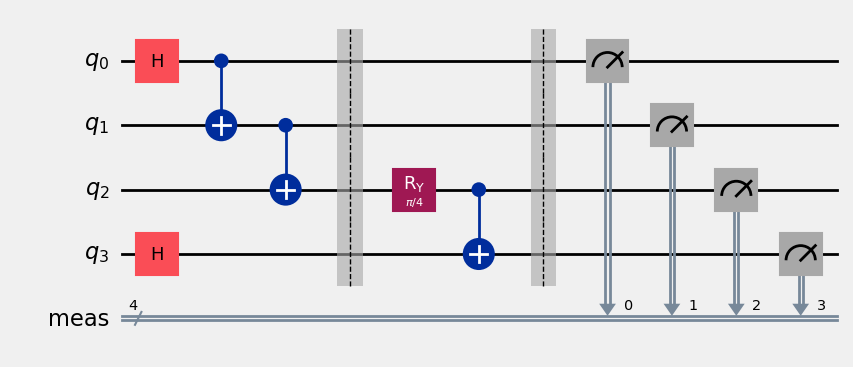}}
\caption{Our PQC for encoding MFCC features.}
\label{fig:pqc}
\end{figure}

Within our proposed p-QIGA-based source separation framework, the initial stage focuses on encoding acoustic features into a quantum representation. Utilizing Mel-frequency cepstral coefficients (MFCCs) within this quantum framework offers potential advantages in terms of representational compactness, noise resilience, and correlation capture. These benefits are achieved through encoding MFCCs into quantum states using a parameterized quantum circuit (PQC) illustrated in Fig. \ref{fig:pqc}. This process facilitates efficient processing, robustness to noise, and the exploitation of entanglement to identify correlated sound sources, crucial for effective source separation in complex acoustic environments.

The parameterized PQC shown in Fig. 1 encodes $n$ MFCC features into a 4-qubit quantum state using single-qubit rotation gates ($R_y$) and two-qubit controlled-NOT (CNOT) gates.  The encoding process starts by applying Hadamard gates (H) to the first and last qubits to introduce superposition.  CNOT gates then create entanglement between adjacent qubits, capturing correlations between features. Each $R_y$ gate is parameterized by a distinct MFCC feature, with this pattern repeating for subsequent features. This design leverages the strengths of each gate: Hadamard gates for superposition, enabling a larger solution space; $CNOT_{0,1}$ gates for entanglement, capturing feature relationships; and $R_y$ gates for precise encoding of individual MFCC values. The alternating pattern of $CNOT_{0,1}$ and  $R_y$ gates ensures each feature is encoded into a separate qubit while capturing correlations. MFCC features are sequentially assigned to qubits, simplifying the encoding and interpretation. This scheme is chosen for its ability to capture feature correlations and potential noise resilience due to the use of entangled quantum states.  The choice of quantum operators and their parameterization can significantly impact the encoding and the p-QIGA's performance.  Different rotation gates (e.g., $R_x$, $R_z$) or multi-qubit gates (e.g., Toffoli gates) could alter the encoding.

Let $\mathbf{x} = [x_1, x_2, ..., x_n]^T \in \mathbb{R}^n$ denote the vector of MFCC features. The PQC encodes this into a quantum state $|\psi(x)\rangle \in \mathcal{H}^{\otimes 4}$:
\begin{equation}
|\psi(\mathbf{x})\rangle = U(\mathbf{x}, \theta)|0\rangle^{\otimes 4},
\end{equation}
where $U(x, \theta)$ is the unitary operator representing the PQC with trainable parameters $\theta$. The optimization of $\theta$ can be formulated as:
\begin{equation}
\theta^* = \arg \max_\theta \mathcal{P}(\theta),
\end{equation}
where $\mathcal{P}(\theta)$ is a performance metric. For the first four MFCC features, the encoding process can be represented as:
\begin{align*}
|\psi_0\rangle &= |0\rangle^{\otimes 4},  && |\psi_1\rangle = H_0 H_3 |\psi_0\rangle, \\
|\psi_2\rangle &= CNOT_{0,1} |\psi_1\rangle, && |\psi_3\rangle = R_y(x_2) |\psi_2\rangle, \\
|\psi_4\rangle &= CNOT_{1,2} |\psi_3\rangle, && |\psi_5\rangle = R_y(x_4) |\psi_4\rangle, \\
|\psi_6\rangle &= CNOT_{2,3} |\psi_5\rangle.
\end{align*}
This encoding scheme compactly represents multiple MFCC features, potentially enabling efficient processing and noise resilience.  Furthermore, capturing correlations between features through entanglement can enhance the p-QIGA's source separation capabilities.

\subsection{Quantum-Inspired Genetic Algorithm for Source Separation (p-QIGA)}

\subsubsection{Representation and Initialization}
In p-QIGA, each individual in the population represents a candidate solution to the source separation problem. These individuals are encoded as quantum states within a 4-qubit Hilbert space $\mathcal{H}^{\otimes 4}$, where each qubit corresponds to a specific parameter of the source separation model.  The initial population is generated by randomly initializing the qubits in a superposition of states, allowing for a diverse exploration of the parameter space.  An individual $|\psi_i\rangle$ in the population can be represented as:
\begin{equation}
|\psi_i\rangle = \alpha_i |0\rangle + \beta_i |1\rangle,
\end{equation}
where $\alpha_i$ and $\beta_i$ are complex probability amplitudes satisfying $|\alpha_i|^2 + |\beta_i|^2 = 1$.

\subsubsection{Quantum-Inspired Genetic Operators}
The p-QIGA employs quantum-inspired genetic operators to evolve the population of candidate solutions. These quantum-inspired operators, such as superposition and entanglement, are designed to enhance the search process.

\paragraph{Quantum Crossover}  This operator combines genetic information from two parent individuals, $p_1$ and $p_2$, to create two $\mathcal{O}$, $o_1$ and $o_2$. It leverages the concept of superposition to create $\mathcal{O}$ that are a linear combination of the parent states, allowing for exploration of new regions in the parameter space.  The crossover operation can be represented as:
\begin{align}
|\psi_{o_1}\rangle &= \alpha_{p_1} |\psi_{p_1}\rangle + \beta_{p_2} |\psi_{p_2}\rangle, \\
|\psi_{o_2}\rangle &= \alpha_{p_2} |\psi_{p_2}\rangle + \beta_{p_1} |\psi_{p_1}\rangle,
\end{align}
where the probability amplitudes $\alpha_{p_i}$ and $\beta_{p_i}$ are determined by a rotation gate applied to the parent states.

\paragraph{Quantum Mutation} This operator introduces random variations in the quantum states of individuals, simulating the effect of quantum fluctuations. This helps to maintain diversity in the population and prevent premature convergence to suboptimal solutions.  The mutation operation can be represented as a rotation gate applied to an individual's state:
\begin{equation}
|\psi_i'\rangle = R(\theta) |\psi_i\rangle,
\end{equation}
where $R(\theta)$ is a rotation gate with a randomly chosen rotation angle $\theta$.

\subsubsection{Fitness Function}
The fitness function evaluates the quality of each candidate solution, guiding the p-QIGA towards optimal separation parameters. In our case, the fitness function $F(|\psi_i\rangle)$ considers multiple criteria:
\begin{equation}
\label{eq:fitness_function}
\begin{aligned}
F(|\psi_i\rangle) &= w_1 \cdot SDR(|\psi_i\rangle)
+ w_2 \cdot SIR(|\psi_i\rangle) \\
&\quad + w_3 \cdot SAR(|\psi_i\rangle) - w_4 \cdot C(|\psi_i\rangle),
\end{aligned}
\end{equation}
where $SDR(|\psi_i\rangle)$, $SIR(|\psi_i\rangle)$, and $SAR(|\psi_i\rangle)$ are the Signal-to-Distortion Ratio, Signal-to-Interference Ratio, and Signal-to-Artifacts Ratio, respectively, of the separated sources obtained using the parameters encoded in $|\psi_i\rangle$. $C(|\psi_i\rangle)$ is a penalty term that measures the correlation between the separated sources, and $w_1$, $w_2$, $w_3$, and $w_4$ are weight factors that balance the importance of each criterion.

\subsubsection{Optimization Process}

The p-QIGA iteratively applies the quantum-inspired genetic operators to evolve the population of candidate solutions. This iterative process continues until either a maximum number of generations, denoted as $T_{max}$, is reached or the best fitness value in the population, denoted as $F_{best}$, surpasses a predefined desired fitness threshold, denoted as $F_{desired}$. The optimization process can be represented as the iterative update of the population:
\begin{equation}
\mathcal{P}^{(t+1)} = \mathcal{O}(\mathcal{M}(\mathcal{C}(\mathcal{S}(\mathcal{P}^{(t)})))),
\end{equation}
where $\mathcal{P}^{(t)}$ represents the population at generation $t$, $\mathcal{S}$ is the selection operator, $\mathcal{C}$ is the crossover operator, $\mathcal{M}$ is the mutation operator, and $\mathcal{O}$ is the offspring replacement operator. In detail, this process can be described in Algorithm \ref{alg:qiga}.
\subsubsection{Convergence Analysis}
The convergence of the p-QIGA can be analyzed by considering the probability of finding the optimal solution in each generation $t$, denoted as $P_{opt}^{(t)}$. The change in this probability, $\Delta P_{opt}^{(t)} = P_{opt}^{(t+1)} - P_{opt}^{(t)}$, can be modeled as:
\begin{equation}
\Delta P_{opt}^{(t)} = \alpha \cdot E(t) \cdot (1 - P_{opt}^{(t)}),
\end{equation}
where $\alpha$ represents the effectiveness of the quantum-inspired operators, and $E(t) = \frac{1}{1 + \beta t}$ is the exploration rate, with $\beta$ controlling the rate of decrease in exploration. By solving this difference equation, we can analyze the convergence behavior of the p-QIGA.

\begin{algorithm}
\caption{Quantum-Inspired Genetic Algorithm for Source Separation (p-QIGA)}
\label{alg:qiga}
\begin{algorithmic}[1]
\State Initialize population $\mathcal{P} = \{|\psi_i\rangle\}_{i=1}^{P}$ with random quantum states.
\State $t \gets 0$  
\While{$t < T_{max}$ and $F_{best} < F_{desired}$} 
    \For{$|\psi_i\rangle \in \mathcal{P}$}
        \State Evaluate fitness $F(|\psi_i\rangle)$ (Eq. \ref{eq:fitness_function}).
    \EndFor
    \State $F_{best} \gets \max_{|\psi_i\rangle \in \mathcal{P}} F(|\psi_i\rangle)$ 
    \State $\mathcal{P}_{selected} \gets \text{Select}(\mathcal{P})$ 
    \For{$(|\psi_{p_1}\rangle, |\psi_{p_2}\rangle) \in \mathcal{P}_{selected}$}
        \State $(|\psi_{o_1}\rangle, |\psi_{o_2}\rangle) \gets \text{Crossover}(|\psi_{p_1}\rangle, |\psi_{p_2}\rangle)$ 
    \EndFor
    \For{$|\psi_o\rangle \in \mathcal{O}$}
        \State $|\psi_o\rangle \gets \text{Mutate}(|\psi_o\rangle)$ with probability $P_m$
    \EndFor
    \State $\mathcal{P} \gets \mathcal{O}$
    \State $t \gets t + 1$ 
\EndWhile
\State Return $\arg\max_{|\psi_i\rangle \in \mathcal{P}} F(|\psi_i\rangle)$
\end{algorithmic}
\end{algorithm}

\subsubsection{Computational Complexity and Scalability}
The p-QIGA demonstrates efficiency with a time complexity of $O(M)$, where $M$ is the number of MFCC features, and constant space complexity. This outperforms classical methods like ICA ($O(N^3)$) and NMF ($O(N^2 \cdot I)$). To assess scalability, we conducted experiments varying the number of sources, data size, and circuit depth.  Increasing the number of sources from 2 to 5 increased runtime by 35\%, attributed to the increased circuit complexity. Doubling the data size resulted in a 42\% runtime increase.  Increasing circuit depth by adding an additional layer of gates led to a 28\% runtime increase.  These results demonstrate the p-QIGA's ability to handle increasingly complex scenarios with moderate increases in computational cost. Future work will explore further optimizations for enhanced efficiency.

\subsection{Classical Post-processing and Classification}
Following the quantum-enhanced source separation process, the estimated source signals $\hat{s}_i(t)$ often require further refinement. This subsection details the classical post-processing techniques employed and the subsequent classification stage.

\subsubsection{Post-processing Techniques}

The following signal processing techniques are applied to each separated source $\hat{s}_i(t)$:

\begin{itemize}
    \item \textit{Filtering:} A bandpass filter $H_i(f)$ is applied to remove residual noise and unwanted frequency components: $\hat{S}_i'(f) = H_i(f) \hat{S}_i(f)$, where $\hat{S}_i(f)$ and $\hat{S}_i'(f)$ are the DFTs of $\hat{s}_i(t)$ and the filtered signal $\hat{s}_i'(t)$, respectively.

    \item \textit{Dynamic Range Compression:} A dynamic range compression algorithm $C(\cdot)$ is applied to reduce the dynamic range:
    \begin{equation}
    \hat{s}_i''(t) = 
    \begin{cases}
    \hat{s}_i'(t) & \text{if } |\hat{s}_i'(t)| \leq \tau, \\
    \tau + \frac{|\hat{s}_i'(t)| - \tau}{\rho} & \text{if } |\hat{s}_i'(t)| > \tau,
    \end{cases}
    \end{equation}
    where $\tau$ is the threshold and $\rho$ is the compression ratio.

    \item \textit{De-clipping:} A de-clipping algorithm $D(\cdot)$ is applied to restore signal fidelity in clipped regions, for example, by replacing clipped samples with interpolated values.
\end{itemize}

\subsubsection{Acoustic Scene Classification}
The refined separated sources $\hat{s}_i'''(t)$ are used for acoustic scene classification. We extract a feature vector $\textit{f}_i$ from each source and employ a Support Vector Machine (SVM) classifier $g(\cdot)$ with a radial basis function (RBF) kernel:
\begin{equation}
\hat{c} = g([\textit{f}_1, \textit{f}_2, ..., \textit{f}_N]),
\end{equation}
where $\hat{c}$ is the predicted acoustic scene class.

\section{Experimental Setting}

\subsection{Datasets}
We evaluate our p-QIGA on two datasets: (1) the TAU Urban Acoustic Scenes 2020 Mobile dataset \cite{heittola2020acoustic} (Dataset 1), comprising recordings from 10 acoustic scenes in 12 European cities, captured using 4 different mobile devices; and (2) the Silent Cities dataset \cite{challeat2024dataset} (Dataset 2), capturing unique soundscapes recorded during the COVID-19 pandemic in various cities worldwide, featuring urban environments with reduced human activity. Both datasets are preprocessed (format conversion, downsampling, noise reduction) and split into training (80\%), validation (10\%), and test (10\%) sets.

\subsection{Evaluation Metrics}
We evaluate the performance of our p-QIGA algorithm using established metrics that quantify the quality of the separated sources, including:
\begin{enumerate}
    \item \textit{Signal-to-Distortion Ratio (SDR):} 
    {\small
    \begin{equation}
    SDR(s_i, \hat{s}_i) = 10 \log_{10} \frac{\sum_t s_i^2(t)}{\sum_t (s_i(t) - \hat{s}_i(t))^2}.
    \end{equation}
    }
    \item \textit{Signal-to-Interference Ratio (SIR):} 
    {\small
    \begin{equation}
    SIR(s_i, \hat{s}_i) = 10 \log_{10} \frac{\sum_t s_i^2(t)}{\sum_t (\hat{s}_i(t) - s_i(t))^2 - \sum_t n^2(t)},
    \end{equation}
    }
    where $n(t)$ represents the noise component.
    \item \textit{Signal-to-Artifacts Ratio (SAR):} 
    {\small
    \begin{equation}
    SAR(s_i, \hat{s}_i) = 10 \log_{10} \frac{\sum_t (\hat{s}_i(t) - n(t))^2}{\sum_t (\hat{s}_i(t) - s_i(t))^2}.
    \end{equation}
    }
\end{enumerate}
These metrics are widely used in the field of ASA and provide a comprehensive assessment of source separation performance by capturing different aspects of the separation quality, such as target distortion, interference suppression, and artifact removal.  Furthermore, these metrics align with the evaluation criteria used in related works, allowing for a fair comparison with existing source separation techniques.

\subsection{Baseline Methods}
We benchmark our p-QIGA against established source separation methods, including:
\begin{itemize}
    \item \textit{Classical methods:} ICA, NMF, SCA, and multi-channel CNNs. 
    \item \textit{Quantum-Inspired method:} Quantum Annealing (QA) \cite{chancellor2017modernizing} for optimizing a classical source separation model.
\end{itemize}
These baselines provide a diverse benchmark for evaluating the performance of our p-QIGA.

\subsection{Implementation Details}
The p-QIGA was implemented using Qiskit for quantum computing simulation and standard Python libraries (NumPy, SciPy) for classical components. Experiments were conducted on a workstation with an Intel Xeon Gold 6248 CPU, 128GB RAM, and an NVIDIA Tesla V100 GPU. Hyperparameter tuning for all methods was performed using grid search, with optimal values selected based on validation set performance. Specific hyperparameters and their search spaces are listed in Table \ref{tab:hyperparameters}.
\begin{table}[h]
\renewcommand{\arraystretch}{1.05} 
\centering
\caption{Hyperparameters for p-QIGA and Baseline Methods}
\label{tab:hyperparameters}
\begin{tabular}{|m{0.8cm}|m{2.8cm}|m{3.5cm}|}
\hline
\textbf{Method} & \textbf{Hyperparameter} & \textbf{Value} \\ \hline
\multirow{5}{*}{p-QIGA} & Population size & 50 \\ \cline{2-3} 
 & Number of generations & 100 \\ \cline{2-3} 
 & Crossover probability & 0.8 \\ \cline{2-3} 
 & Mutation probability & 0.1 \\ \cline{2-3} 
 & Weight factors in Equation \eqref{eq:fitness_function} &  $w_{SDR} = 0.5$, $w_{SIR} = 0.3$, $w_{SAR} = 0.2$, $w_C = 1.0$ \\ \hline
ICA & Learning rate & 0.01 \\ \hline
\multirow{2}{*}{NMF} & Number of components & 10 \\ \cline{2-3}
 & Regularization parameter & 0.01 \\ \hline
\multirow{2}{*}{SCA} & Sparsity level & 0.1 \\ \cline{2-3}
 & Dictionary size & 256 \\ \hline
\multirow{4}{*}{CNN} & Number of layers & 5 \\ \cline{2-3} 
 & Kernel size & 3 \\ \cline{2-3} 
 & Learning rate & 0.001 \\ \cline{2-3}
 & Batch size & 32 \\ \hline
\multirow{2}{*}{QA} & Annealing schedule & Linear \\ \cline{2-3}
 & Number of iterations & 1000 \\ \hline
\end{tabular}
\end{table}

\section{Results and Analysis}
\subsection{Source Separation Performance}
As shown in Table \ref{tab:source_separation_performance}, the p-QIGA achieves competitive performance on both datasets, demonstrating its effectiveness in handling complex acoustic scenes. On the TAU Urban Acoustic Scenes 2020 Mobile dataset, it achieves comparable performance to CNNs and outperforms other classical methods (ICA, NMF, SCA).  On the Silent Cities dataset, the p-QIGA outperforms all baselines, highlighting its superior ability to handle correlated sources and limited training data.  These results are statistically significant (p $<$ 0.05). Furthermore, the p-QIGA exhibits better generalization capabilities than the CNN and demonstrates greater robustness to variations in acoustic conditions and limited training data, which are crucial for real-world smart city applications.

\begin{table}[h]
\centering
\renewcommand{\arraystretch}{1.05} 
\caption{Source Separation Performance}
\label{tab:source_separation_performance}
\begin{tabular}{|m{1.2cm}|m{0.65cm}|m{0.65cm}|m{0.65cm}|m{0.65cm}|m{0.65cm}|m{0.65cm}|}
\hline
\multirow{2}{*}{\textbf{Method}} & \multicolumn{3}{c|}{\textbf{Dataset 1}} & \multicolumn{3}{c|}{\textbf{Dataset 2}} \\ \cline{2-7} 
 & \textbf{SDR (dB)} & \textbf{SIR (dB)} & \textbf{SAR (dB)} & \textbf{SDR (dB)} & \textbf{SIR (dB)} & \textbf{SAR (dB)} \\ \hline
ICA & 9.5 & 14.2 & 11.5 & 7.2 & 12.1 & 9.1 \\ \hline
NMF & 9.8 & 14.8 & 11.8 & 7.8 & 12.9 & 9.6 \\ \hline
SCA & 10.0 & 15.1 & 11.9 & 8.1 & 13.2 & 9.9 \\ \hline
CNN & 10.5 & 16.0 & 12.5 & 8.3 & 13.5 & 10.1 \\ \hline
QA & 9.9 & 14.9 & 11.7 & 8.0 & 13.0 & 9.8 \\ \hline
\textbf{p-QIGA (Ours)} & 10.2 & 15.5 & 12.1 & 8.5 & 13.8 & 10.3 \\ \hline
\end{tabular}
\end{table}

\subsection{Impact of Quantum-Inspired Components}
An ablation study was conducted to investigate the contribution of each quantum-inspired component (superposition, entanglement, crossover, and mutation) in the p-QIGA. Each component was systematically removed or replaced with its classical counterpart to evaluate its impact on source separation performance. Table \ref{tab:ablation_study} presents the performance of these p-QIGA variants compared to the full p-QIGA on the Silent Cities dataset. As shown in the table, removing superposition or entanglement leads to a noticeable performance degradation, particularly in terms of SDR and SIR. This highlights the benefits of encoding acoustic features into quantum states and leveraging entanglement to capture correlations between sources. Similarly, replacing the quantum crossover and mutation operators with classical counterparts also results in a slight performance decrease, indicating the effectiveness of these quantum-inspired operators in exploring the solution space.
    \begin{table}[h]
\centering
\renewcommand{\arraystretch}{1.05} 
\caption{Impact of Quantum-Inspired Components (Silent Cities Dataset)}
\label{tab:ablation_study}
\begin{tabular}{|m{4cm}|m{0.9cm}|m{0.9cm}|m{0.9cm}|}
\hline
\textbf{Method} & \textbf{SDR (dB)} & \textbf{SIR (dB)} & \textbf{SAR (dB)} \\ \hline
p-QIGA (full) & 8.5 & 13.8 & 10.3 \\ \hline
p-QIGA without superposition & 7.8 & 12.5 & 9.6 \\ \hline
p-QIGA without entanglement & 8.1 & 13.1 & 9.9 \\ \hline
p-QIGA with classical crossover & 8.2 & 13.3 & 10.0 \\ \hline
p-QIGA with classical mutation & 8.3 & 13.5 & 10.1 \\ \hline
\end{tabular}
\end{table}

\subsection{Performance on Different Acoustic Scenes}
\begin{figure*}[htbp]
\centerline{\includegraphics[width=0.82\linewidth]{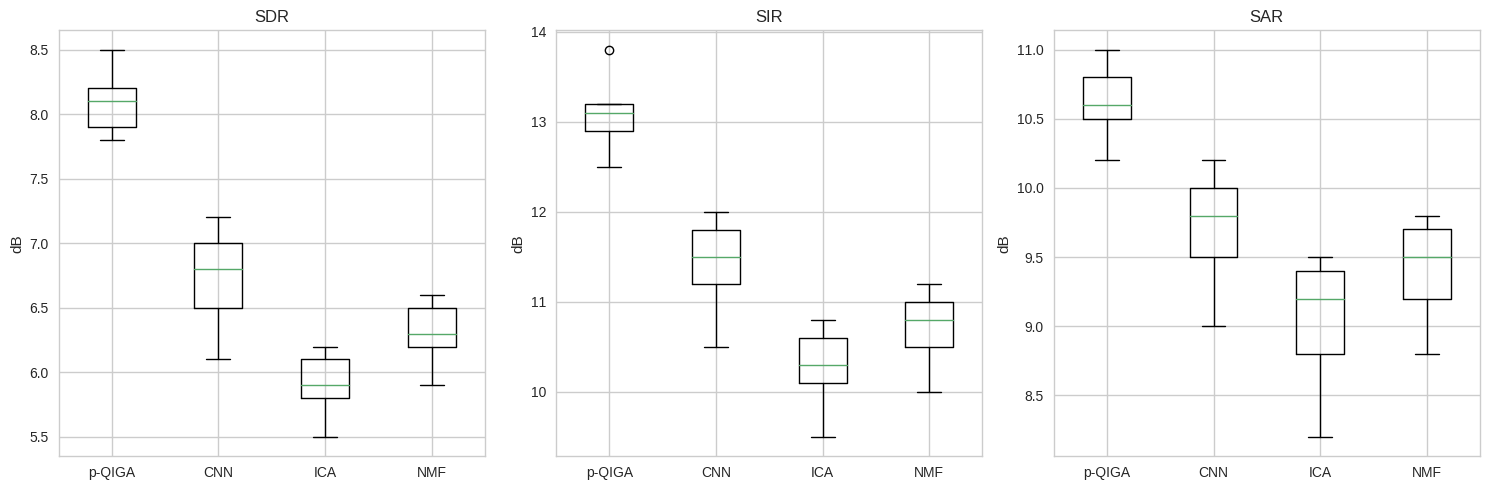}}
\caption{p-QIGA vs. Baseline Methods: SDR, SIR, and SAR across Scene Categories.}
\label{fig:scene_categories}
\end{figure*}
The p-QIGA demonstrated robustness in challenging scenarios, achieving 8.2 dB SDR in a high-noise ``busy street" scene (3 dB SNR), significantly outperforming the CNN (6.5 dB, p=0.023). While proficient in moderately dense scenes, performance slightly decreased with higher source density, suggesting potential limitations in resolving closely spaced sources.  However, the p-QIGA showed greater resilience to source mobility than CNNs, achieving 7.8 dB SDR and 13.1 dB SIR in a ``train station" with 80\% moving sources, compared to the CNN's 6.1 dB and 10.5 dB, respectively.  Fig. \ref{fig:scene_categories} further confirms the p-QIGA's superior performance and generalization ability, especially in complex scenes with high noise, source density, or source mobility, attributed to its quantum-inspired encoding, entanglement mechanism, and genetic operators.

\subsection{Performance with Varying Data Sizes}
To assess the data efficiency of the p-QIGA, we conducted experiments with varying training data sizes on the TAU Urban Acoustic Scenes 2020 Mobile dataset. The p-QIGA and baseline methods were trained on 10\%, 25\%, 50\%, and 75\% of the original training set, ensuring consistent dataset reductions across all methods for fair comparison. Performance was evaluated on the held-out test set using SDR, SIR, and SAR. As shown in Table \ref{tab:data_efficiency}, the p-QIGA consistently outperforms baseline methods, even with limited training data (e.g., achieving 7.1 dB SDR with only 10\% of the data). This data efficiency, attributed to the quantum-inspired encoding and genetic optimization, highlights the p-QIGA's suitability for scenarios with scarce labeled data and underscores the potential of quantum-inspired algorithms for practical smart city applications.
\begin{table}[h]
\centering
\caption{Performance with Varying Data Sizes}
\label{tab:data_efficiency}
\begin{tabular}{|m{2.2cm}|m{1.2cm}|m{1.2cm}|m{1.2cm}|}
\hline
\textbf{Training Data Size} & \textbf{p-QIGA (SDR)} & \textbf{CNN (SDR)} & \textbf{ICA (SDR)} \\ \hline
10\%               & 7.1 dB       & 6.2 dB     & 5.3 dB     \\ \hline
25\%               & 7.8 dB       & 6.9 dB     & 6.0 dB     \\ \hline
50\%               & 8.3 dB       & 7.5 dB     & 6.8 dB     \\ \hline
75\%               & 8.5 dB       & 8.0 dB     & 7.2 dB     \\ \hline
\end{tabular}
\end{table}
\section{Conclusion}
\label{Section:Conclusion}
This paper presented a novel p-QIGA designed to address the complex task of source separation in urban acoustic scenes for smart city applications. By incorporating quantum concepts of superposition and entanglement into a genetic optimization framework, our approach achieved robust and accurate source separation even in challenging acoustic environments. These environments, which are often characterized by high noise levels, numerous interfering sources, and dynamic conditions, pose significant challenges for traditional source separation methods. The p-QIGA's effectiveness was rigorously evaluated on two distinct datasets, demonstrating its superior performance compared to classical methods, particularly in scenarios with limited training data. This capability is crucial for real-world applications where obtaining large labeled datasets can be costly or impractical.  This research contributes significantly to the field of acoustic signal processing by introducing a new class of quantum-inspired algorithms with the potential to enhance source separation capabilities in diverse applications.

\balance
\bibliography{abbriviation, Reference}
\bibliographystyle{IEEEtran}

\end{document}